\documentstyle[aps,twocolumn,epsf]{revtex}
\begin{document}
\draft
\title{
Heisenberg Spin-One Chain in  Staggered Magnetic Field : \\
A Density Matrix Renormalization Group Study
}
\author{Jizhong Lou, Xi Dai, Shaojin Qin, Zhaobin Su}

\address{
Institute of Theoretical Physics, P. O. Box 2735, Beijing
100080, P.R. China
}
\author{Lu Yu}
\address{
International Center for Theoretical Physics, Trieste 34100, Italy \\
and Institute of Theoretical Physics, P. O. Box 2735, Beijing
100080, P.R. China
}
\date{ \today }
\maketitle
\begin{abstract}

Using the density matrix renormalization group technique, we
calculate numerically the low energy excitation spectrum
and magnetization curve of
 the  spin-1 antiferromagnetic chain in a
staggered magnetic field, which is expected to describe the
physics of $R_2 Ba Ni O_5 (R \neq Y) $  family below
the N\'eel temperature of the magnetic rare-earth ($R$) sublattice.
These  results  are valid in the entire range of the staggered field,
and agree with those given by the non-linear $\sigma$ model
study for small fields,
but differ from the latter for  large fields.
They are consistent with the available experimental data.
The correlation functions for this model are also calculated.
The  transverse correlations display
the anticipated exponential  decay  with  shorter
correlation length, while the  longitudinal correlations show
explicitly  the induced staggered magnetization.

\end{abstract}

\narrowtext

The quasi-one-dimensional magnets have been the focus of analytic,
numerical and experimental studies since Haldane pointed out the
difference between the integer spin Heisenberg antiferromagnetic (AF)
chains and the half-integer chains in 1983.$^{\cite{Haldane}}$
By mapping the Heisenberg spin chains onto the $O(3)$ nonlinear
$\sigma$-model,$^{\cite{Affleck}}$
he   conjectured that the low-energy excitation spectrum displays
a finite  gap for the integer spin systems while it is gapless for
half-integer spin chains. This conjecture has been verified
 by later experiments on quasi-one-dimensional spin-1 materials
such as NENP  and $Y_2 Ba Ni O_5$  which show clear evidence of the Haldane
gap.$^{\cite{GAP}}$  Nowadays, the pure one-dimensional Haldane systems are
 fairly well understood, and a reliable estimate for the Haldane gap
$\Delta = 0.41048(2) J$ for spin-1 chains has been  obtained  by both density
matrix renormalization group (DMRG) calculation$^{\cite{White}}$ and   finite
size exact diagonalization.$^{\cite{Golinelli}}$

More recent developments on the Haldane systems concern   various
effects of external perturbations: doping with magnetic or
non-magnetic impurities$^{\cite{DopedE}}$ and applying  external magnetic
field. The impurity  doping  may introduce   bound states within the
Haldane gap,$^{\cite{DopedT}}$  while applying
 uniform external magnetic field splits the degenerate Haldane
triplet state into transverse and longitudinal modes.$^{\cite{HighH}}$
The longitudinal mode becomes softened
upon   increase of the magnetic field, and at a critical field $H_c$
the system enters a new phase
with long-range AF order. Of course, a staggered
applied magnetic field is even more interesting
which  would induce  non-vanishing staggered
magnetization and  affect the Haldane gap excitation spectrum,
but such a staggered field
cannot be materialized by an external source.

Most recently, a series of experiments performed on the family of
quasi-one-dimensional materials with a general formula
$R_2 Ba Ni O_5, ^{\cite{Xu,Sa,Zhe1,Zhe2,Zhe3,Yo1,Yo2,Zhe4}}$
where $R$ is one
of the magnetic rare-earth elements substituting fully or partially $Y$
(for brevity we denote this replacement by $R \neq Y$), have made it
possible to study the effect
of  the staggered magnetic field on the Haldane systems in detail. All
members of
this family contain spin-1 $Ni^{2+}$  linear chains and the in-chain
AF exchange coupling is rather strong. (The detailed structure of
this family of compounds is described in Ref.{\cite{Zhe1}}).
The reference compound $Y_2 Ba Ni O_5$
is found to be highly one-dimensional with negligible interchain
interactions,$^{\cite{Xu}}$
 and no magnetic order has been observed so far even at  very low
temperatures.$^{\cite{DopedE}}$  Hence it
is believed to be an almost ideal  example of the Haldane-gap system.
Other members have magnetic $R^{3+}$ ions in addition to the spin-1 $Ni^{2+}$
ions. These ions are positioned between two neighboring $Ni$ chains, weakly
coupled
to the $Ni^{2+}$  ions   and the coupling between  themselves  is also very
weak.
Nevertheless, these magnetic ions are  AF ordered below certain N\'eel
temperature $T_N$.
These ions do not affect the $Ni$ chains substantially above
$T_N$,  keeping   their Haldane features untouched,
but   the 3D  AF ordered $R^{3+}$  sublattice below $T_N$
 has dramatical effects on these chains, imposing  a  staggered
magnetic field. The neutron scattering experiments on
powder samples and small size single-crystals of $Nd_2 Ba Ni O_5$ and
$Pr_2 Ba Ni O_5 ^{\cite{Zhe3,Yo2,Zhe4}}$ show
an increase of the energy gap below the N\'{e}el
temperature.

	We assume that these chains can be still considered one-dimensional,
being  put in a staggered field created by 3D ordered $R^{3+}$ ions
at low temperatures. The Hamiltonian  can be then writen as :

\begin{equation}
H=J\sum_{i} \left[{\bf S}_i \cdot {\bf S}_{i+1}+ h (-1)^i
S^z_i\right],
\label{StagH}
\end{equation}
where $J$ is the exchange constant (to be taken as energy unit,
i.e.,   $J$ =1 ). The dimensionless staggered
field      $h=g S \mu_B H_{\pi}/J$
with  $H_{\pi}$ as  the physical staggered
field, which, in turn,  is proportional to  the $R$ sublattice
magnetization $M_R$
\begin{equation}
H_{\pi}= \alpha M_R.
\label{SHF}
\end{equation}
$g$=2 is the theoretically predicted gyromagnetic ratio of the $Ni$ ion.
The direction of the staggered field
has been  chosen  as the $z$ axis.
This Hamiltonian has been considered using the mean-field
theory$^{\cite{Ma,Zhe4}}$
as well as  by mapping onto the O(3) nonlinear $\sigma$
model (NLSM).$^{\cite{MA2}}$

In this Communication we use the DMRG technique to
calculate the low energy excitation spectrum
and magnetization curve of the  Hamiltonian Eq.(\ref{StagH}).
The obtained field dependence of the gap and the staggered magnetization
is consistent with the experimental results.
 These  results  are valid in the entire range of the staggered field
and recover those given by non-linear $\sigma$ model for small fields
but differ from the latter for large fields.
Moreover, we calculate the spin-spin
correlation functions for this model.
The  transverse correlations display an exponential  decay
as anticipated for the spin-1 AF chain,  with a shorter correlation
length, while the  longitudinal correlations show explicitly  the induced
staggered magnetization.

We   follow  the standard DMRG algorithm$^{\cite{White,DMRG,QWY}}$
to calculate the low-energy excitations of  the Hamiltonian (\ref{StagH}),
adopting the periodic boundary conditions (PBC).
We use the infinite-chain  algorithm
 up to chain length $N$=60 and keep as many as 400 optimized
states during  each
sweep. The largest truncation errors are of the order of  $10^{-8}$ for smaller
$h$, while for bigger $h$, these errors are  as small as $10^{-13}$,
which means our results are even more reliable for bigger $h$.

\begin{figure}[hbt]
\epsfxsize=\columnwidth
\epsfbox{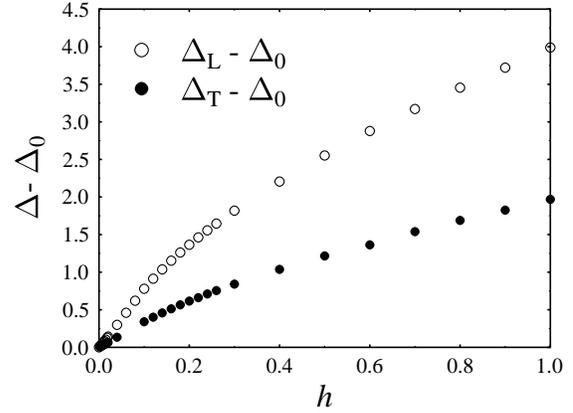}
\caption{ The DMRG results for the   transverse (solid circle) and
longitudinal (empty circle) energy
gaps as functions of
the  staggered magnetic field for  spin-1 chain.
}
\label{GapValue}
\end{figure}

The numerical results for the change of the lowest excitation energies
(the Haldane gap) of  Hamiltonain (\ref{StagH})
$\Delta - \Delta_0$
are presented in Fig.  \ref{GapValue}  as
functions of the dimensionless staggered magnetic
field $h$. In the absence of this field the longitudinal ($\Delta_L$) and
transverse ($\Delta_T$) modes are degenerate,
forming the Haldane triplet. For
non-zero $h$, these  modes will split with respect to each other. Both
of them will increase with the staggered magnetic field, while
the longitudinal gap increases   faster than the transverse one.
For small staggered fields, the increase of the longitudinal gap will
be nearly three times faster then the transverse ones, while for   larger
staggered fields, this ratio will  decrease,  and is is approximately two
for the largest staggered field we considered.

\begin{figure}[hbt]
\epsfxsize=\columnwidth
\epsfbox{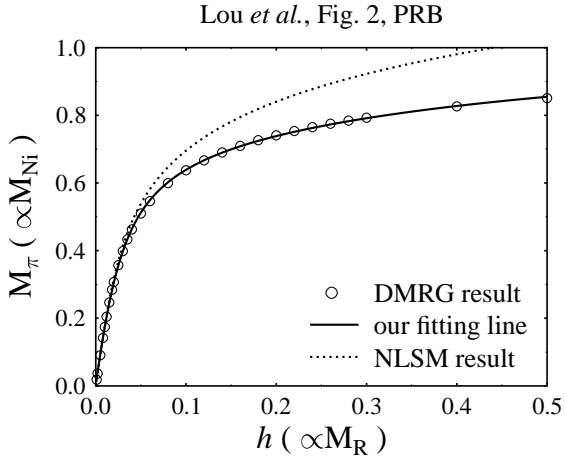}
\caption{
The staggered magnetization  curve for spin-1 chain.
The staggered magnetic field   is proportional to
the magnetization    of the $R$ sublattice. The numerical
results (solid circle)   are fitted by a function with four
parameters (solid line) (see Eq. (\ref{MOM} )),
with $a$=0.412, $b$=38.106, $c$=1.195, $d$= 0.621, respectively; the
analytic results given  by non-linear $\sigma$ model are
also presented (dotted line).
}
\label{MOMENT}
\end{figure}

The staggered magnetic moment of the system with chain length $N$
is defined as
\begin{equation}
M_{\pi} ( N )=\frac{1}{N} \sum_{i} (-1)^{i} \langle S_{i}^{z} \rangle,
\end{equation}
the longest chain in our calculation being N=60.
Then the staggered magnetic moment for infinite long chain $M_\pi$ can be
obtained by extrapolating
$M_{\pi} = \lim_{N \rightarrow \infty} M_{\pi} (N)$.
Obviously, this quantity is a function of the
staggered field, and our numerical results are shown in Fig. \ref{MOMENT}.
Considering Eq. (\ref{SHF}), this figure is nothing but   the relation
between the magnetization of the $Ni$ sublattice and the $R$
sublattice, and it  is qualitatively in agreement with the
exprimental data in Fig. \ref{MOMENT} of Ref.
\cite{Zhe4}.
Our numerical results show  that  in small staggered fields, the magnetization
change linearly with the increase of the field,  so we can easily extract the
"zero field" staggered magnetic susceptibility
$\chi^{(s)} (0)$=18.50/J. This  value fully  agrees with the
results obtained from the transfer-matrix  renormalizaion
group $^{\cite{TMRG}}$, Quantum Monte Carlo $^{\cite{MentoCarlo}}$ and the
NLSM calculations.$^{\cite{MA2}}$ We fit our results using the following
function:
\begin{equation}
M_\pi =
a~ \arctan (b~ {\it h}) + (1-\frac{\pi}{2} a) \tanh (c~ {\it h}^d)
\label{MOM}
\end{equation}
with $a$=0.412,
$b$=38.106,
$c$=1.195,
$d$=0.621. The fittig line is  also shown
in Fig. \ref{MOMENT}.
To compare with the NLSM results $^{\cite{MA2}}$ in detail, in Fig.
\ref{MOMENT},
we use their analytic relation
$$ \chi^{(s)} (0) h = M_\pi ( 1 + 1.56 M_\pi^2 + 2.4 M_\pi^4 + 3.27 M_\pi^6) $$
with their value $\chi^{(s)} (0)$ = 18.7/J as a reference. We   see clearly
that the analytic expression  is very good  for small staggered fields,
while for
larger staggered fields it deviates from our numerical
results significantly.
We have also calculated the magnetic moment for large $h$ which is
not shown in Fig. \ref{MOMENT}. Our result indicates that the
moment  should  saturate in large enough staggered
magnetic field  for both isotropic  and (single-ion) anisotropic cases.
The unsaturated moment at zero-temperature observed so far in
various experiments tells us that the induced staggered magnetic
field on $Ni$ chains is, probably,   not large enough yet. This issue was
also discussed earlier.$^{\cite{Zhe4}}$

\begin{figure}[hbt]
\epsfxsize=\columnwidth
\epsfbox{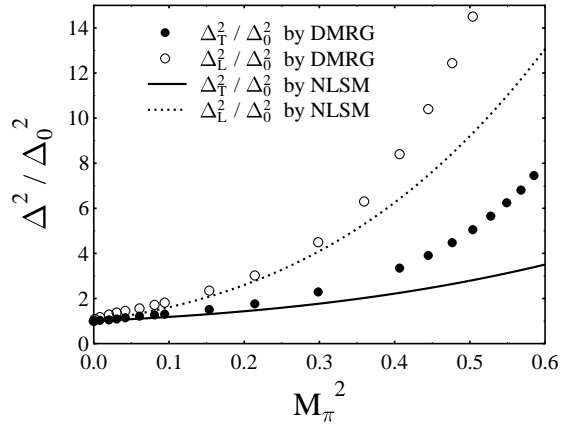}
\caption{
The numerical results for the transverse gaps (solid circle) and
the longitudinal gaps (empty circle)  {\it versus} the magnetic moment
on $Ni$ sites; the NLSM results are shown
 ( solid line for transverse gaps and  dotted line for longitudinal
gaps) in the same figure.
}
\label{MO-GAP}
\end{figure}

  From the above results, we  obtain  the values of  the transverse
as well as  the longitudinal gap as functions of the magnetic
 moment of the $Ni$ sites, which
 can be compared directly with the analytic NLSM result
 (see Fig. \ref{MO-GAP}).  These results are also consistent with
the experimental data.$^{\cite{Zhe3,MA2}}$
As for the comparison with the NLSM treatment
 we  see again that in small staggered fields, the analytic
and  numerical results
are in  good agreement with each other, while for larger staggered fields,
the analytic results deviate significantly from the numerical simulations.
Both longitudinal and transverse  gaps increase  faster in simulations
than the  NLSM predicts.
Since for our DMRG  calculations, larger is  the staggered field,
more reliable are the results, so   the disagreement of magnetization moment
and the gaps between these two approaches raises a question
whether  the NLSM mapping is valid or not
for large staggered magnetic fields.

\begin{figure}[hbt]
\epsfxsize=\columnwidth
\epsfbox{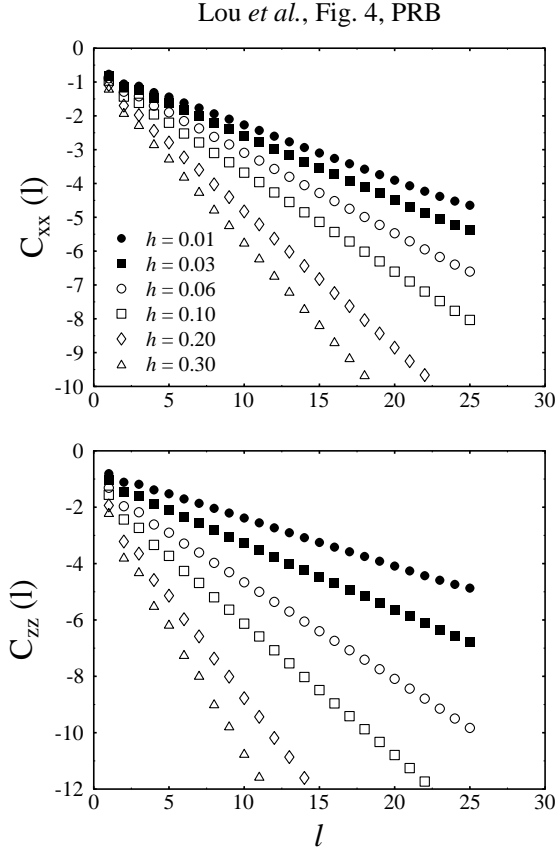}
\caption{Spin-spin correlation functions for different values of staggered
magnetic field,
where $l$ is  the distance between  the two spins.
Both correlations $\langle S^x_0 S^x_i \rangle$  and
$\langle S^z_0 S^z_i \rangle - \langle S^z_0 \rangle \langle S^z_i \rangle$
decay  exponentially, but the latter decays faster.}
\label{CORRELATION}
\end{figure}

\begin{figure}[hbt]
\epsfxsize=\columnwidth
\epsfbox{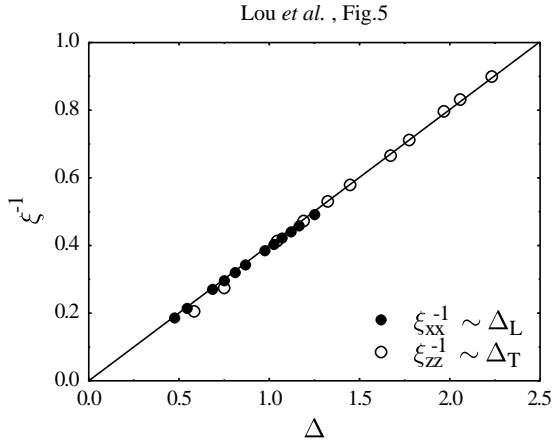}
\caption{ The inverse   correlation lengths $\xi^{-1}_{xx}$
and $\xi^{-1}_{zz}$ for infinite chain length vs.
the transverse   and  longitudinal gaps,  respectively.  The solid
line is the fiting   $\xi^{-1}=0.402*\Delta$. }
\label{SPINWAVE}
\end{figure}

Besides calculating the low-energy spectrum of the one-dimensional systems,
the DMRG method also provides a direct and simple
way to calculate the spin-spin correlation functions
which can shed some further light on the
nature of the system under study.  For a pure Haldane system, the
correlation dacays
exponentially, following
\begin{equation}
\langle S^{x(z)}_0 S^{x(z)}_l \rangle =
(-1)^l A \frac{e^{-\frac{l}{\xi}}}{\sqrt{l}},
\label{PureC}
\end{equation}
where $\xi$=6.03 is the correlation length
obtained by the numerical study,$^{\cite{White}}$  and A is a constant.
When  a staggered field is applied, the AF long-range  order will be induced
along the $z$-direction,  so $\langle S^z_0 S^z_l \rangle $ will not decay
any more.
However,  $\langle S^z_0 S^z_l \rangle -
\langle S^z_0 \rangle \langle S^z_l \rangle$ will still decay exponentially.
Of course, the transverse correlations $\langle S^x_0 S^x_l \rangle$
decay exponentially as before,  but  with    modified exponents.
In Fig. \ref{CORRELATION}(a) and
Fig. \ref{CORRELATION}(b), the functions
$C_{xx}(l)=\ln (\sqrt{l} |\langle S^x_0 S^x_l \rangle|)$  and
$C_{zz}(l)=\ln (\sqrt{l} |\langle S^z_0 S^z_l \rangle
-\langle S^z_0 \rangle \langle S^z_l \rangle|)$
are shown for different staggered
fields. We   see that both  functions decay
exponentially following Eq. (\ref{PureC})
and the correlation lengths $\xi_{xx}$ and $\xi_{zz}$ decrease
with increasing staggered field.
As $h$ increases, the reduced longitudinal correlations decay much
faster.
We have extracted the correlation lengths and extrapolated them
to  infinite chain length
by considering the chain length dependence of $\xi_{xx}$ and $\xi_{zz}$.
We find that $\xi^{-1}_{xx} \sim \Delta_T$  and $\xi^{-1}_{zz} \sim
\Delta_L$,
and both  results can be
fitted by $\xi^{-1} = 0.402*\Delta$ (Fig. \ref{SPINWAVE}),
which coincides exactly
with  $\xi = 6.03 $ obtained for the isotropic spin chain
($h=0$, $\Delta_0= 0.41$).$^{\cite{White}}$  This is an independent check
of the self-consistency in our calculations.

In conclusion, by considering a model hamiltonian which
describes the physics of a family of mixed-spin materials in
the temperature range below $T_N$ of the magnetic rare-earth sublattice,
we have calculated numerically
the energy gap and the staggered magnetic moment as functions of the
 staggered magnetic field created by the   AF long range order.
The obtained results are consistent  with
the experimental data qualitatively.  Our numerical results are also
compared with the analytic considerations based on the non-linear
$\sigma$ model.  The comparison shows that the NLSM results
are good for small staggered fields, while they
deviate from the numerical simulations for larger staggered fields.
After submitting this paper we saw a  new report on polarized neutron study of
longitudinal Haldane-gap excitations in Nd$_2$BaNiO$_5$.$^{\cite{Ray}}$
which show somewhat different behavior than expected from the theory.
The reason of this discrepancy has to be understood.

Our work was supported by the  National Natural Science Foundation
of China (NFSC).
The computations were performed on machines in the State Key Laboratory of
Science and Engeneering of China (LSEC) and the Chinese Center of Advanced
Science and Technology (CCAST). One of the authors (J. Lou) would like to
thank Prof. T. K. Ng and Dr. Tao Xiang for helpful discussions.

\end{document}